\documentclass[aps,prb,twocolumn,superscriptaddress,showpacs,showkeys,amsmath,amssymb]{revtex4}
\usepackage{epsfig,amsmath,amssymb,color}
\begin{document}

\title{On the compressibility of deformable spin chains in a vicinity of quantum critical points}

\author{Oleg Derzhko}
\affiliation{Institute for Condensed Matter Physics,
             National Academy of Sciences of Ukraine,
             1 Svientsitskii Street, L'viv-11, 79011, Ukraine}
\affiliation{Department for Theoretical Physics,
             Ivan Franko National University of L'viv,
             12 Drahomanov Street, L'viv-5, 79005, Ukraine}
\affiliation{Abdus Salam International Centre for Theoretical Physics,
             Strada Costiera 11, I-34151 Trieste, Italy}

\author{Jozef Stre\v{c}ka}
\affiliation{Department of Theoretical Physics and Astrophysics,
             Faculty of Science, P.~J.~\v{S}af\'{a}rik University,
             Park Angelinum 9, 040 01 Ko\v{s}ice, Slovak Republic}

\author{Lucia G\'{a}lisov\'{a}}
\affiliation{Department of Applied Mathematics and Informatics,
             Faculty of Mechanical Engineering, Technical University,
             Letn\'{a} 9, 042 00 Ko\v{s}ice, Slovak Republic}

\date{\today}

\begin{abstract}
We calculate the ground-state compressibility of a deformable spin-1/2 Heisenberg-Ising chain with Dzyaloshinskii-Moriya interaction 
to discuss how a quantum critical point inherent in this spin system
may manifest itself in the elastic properties of the underlying lattice.
We compare these results with the corresponding ones 
for the spin-1/2 Ising chain in a longitudinal or transverse field and the spin-1/2 $XX$ chain in a transverse field. 
The inverse compressibility of the spin-1/2 $XX$ chain in a transverse field 
exhibits a hysteresis in a vicinity of quantum critical point that is accompanied with the finite jump of compressibility. 
Contrary to this, 
the inverse compressibility diminishes continuously close to a quantum critical point 
of the spin-1/2 Ising chain in a transverse field and the spin-1/2 Heisenberg-Ising bond alternating chain.
\end{abstract}



\maketitle

\section{Introductory remarks}
\label{sec1}
\setcounter{equation}{0}

Quantum spin chains provide an excellent play-ground for theoretical studies of collective quantum phenomena.
They may exhibit very rich low-temperature behavior
showing numerous quantum phases
(sometimes exotic ones which are characterized by non-conventional order parameters)
and quantum critical points\cite{1}.
Moreover, including spin-lattice interactions may lead to new phenomena
like, e.g., spin-Peierls instabilities\cite{2,3}.
Importantly,
quantum spin chains in many cases are amenable to rigorous analysis
that permits to obtain results free of using any uncontrolled approximations\cite{bethe,lieb,katsura,encyclopedia,parkinson}.
At the same time, real solid-state representatives of exactly solved quantum spin-chain models have become available 
due to a recent progress in material sciences and comparison of theoretical predictions with experimental observations 
has thus become possible\cite{3,xy}.

In the present paper we wish to discuss
how quantum critical point,
inherent in a system of interacting quantum spins,
may manifest itself
in the properties of the underlying lattice.
For this purpose we consider simple but nontrivial exactly solvable spin models
focusing in particular on a spin-1/2 Heisenberg-Ising chain with Dzyaloshinskii-Moriya interaction.
We also consider a harmonic one-dimensional lattice
and use an adiabatic treatment assuming that the lattice may rearrange itself
to suit a ``request'' of the interacting quantum spins.
The results to be obtained will be confronted with the corresponding ones for three exactly soluble quantum spin chains, 
more specifically, the spin-1/2 Ising chain in a longitudinal or transverse field and the spin-1/2 $XX$ chain in a transverse field.
In our study we calculate the lattice compressibility
to show how its behavior indicates peculiarities in those quantum spin systems.
Our goal is not to describe any specific material,
but to analyze accurately model systems amenable to exact calculations.
We believe that our analysis of correlation between lattice properties and peculiar behavior of spin system
will have some qualitative merit in general.

The paper is organized as follows.
First, in Sec.~\ref{sec2} we illustrate a calculation scheme 
considering some well-known examples of exactly solved quantum spin chains
including the spin-1/2 Ising chain in a longitudinal field, as well as, the spin-1/2 $XX$ chain in a transverse field.
In Sec.~\ref{sec3} we examine the elastic properties of the deformable spin-1/2 Ising chain in a transverse field.
Our main emphasis is laid on the deformable spin-1/2 Heisenberg-Ising chain with Dzyaloshinskii-Moriya interaction considered in Sec.~\ref{sec4}.
In this section we examine the compressibility of the spin lattice
by discussing its relation to a quantum critical point inherent in the spin system.
Finally, our findings will be summarized in Sec.~\ref{sec5}.

\section{Compressibility of some quantum spin-chain models}
\label{sec2}
\setcounter{equation}{0}

In this section, we discuss a calculation of the compressibility of a lattice with interacting spins.
First, consider an ``empty'' (i.e., without spins) one-dimensional lattice of $N$ sites
which is perfectly periodic and has minimum energy for some lattice constant $a_0$.
Applying pressure along the lattice we may either elongate or shorten it.
We assume that the extension or reduction of the lattice length is uniform,
i.e., the change of distance between any two neighboring sites is in equilibrium $a-a_0\propto\delta$,
and the total elastic energy penalty is $N\alpha\delta^2/2$,
where $\alpha$ is the (bare) elastic constant of the lattice.
We consider the variational Gibbs free energy per site $g(T,p;\delta)$,
or more precisely, the variational enthalpy per site $g(T=0,p;\delta)$
since we restrict ourselves in what follows to the zero-temperature case
\begin{eqnarray}
\label{2.01}
g(T=0,p;\delta)=\frac{1}{2}\alpha\delta^2+p\delta,
\end{eqnarray}
where $p$ is the (dimensionless) pressure.
The equilibrium value of $\delta$ denoted further as $\delta(p)$
is obtained by minimizing the variational enthalpy $g(T=0,p;\delta)$ (\ref{2.01}) 
with respect to $\delta$ (see Refs.~\onlinecite{mattis,orignac}),
i.e., by solving the equation $0=\partial g(T=0,p;\delta)/\partial\delta=\alpha \delta+p$,
with the result
\begin{eqnarray}
\label{2.02}
\delta(p)=-\frac{p}{\alpha}.
\end{eqnarray}
Note that $\delta(p)>0$ ($\delta(p)<0$)
[i.e., stretching (shrinking) of the lattice occurs] 
if $p<0$ ($p>0$).
Substituting $\delta=\delta(p)$ (\ref{2.02}) into Eq. (\ref{2.01})
we obtain the enthalpy per site
$g(T=0,p)=g(T=0,p;\delta(p))=-p^2/(2\alpha)$.
Moreover,
using Eq. (\ref{2.02}) we can calculate the observed lattice compressibility (or the inverse elastic constant)
\begin{eqnarray}
\label{2.03}
\frac{1}{\varkappa}\equiv -\left(\frac{\partial \delta(p)}{\partial p}\right)
=\frac{1}{\alpha}.
\end{eqnarray}
Obviously,
we have found that the observed elastic constant of the lattice $\varkappa$
coincides with the bare elastic constant $\alpha$
as it should for the case at hand.

Now we populate this lattice with $N$ interacting quantum spins $s=1/2$
which are represented by Pauli matrices.
We assume only the nearest-neighbor exchange interactions between the spins
which obviously depend on the intersite distance.
Thus, Eq. (\ref{2.01}) will now contain on the right-hand side a $\delta$-dependent term $e_0(\delta)$ 
corresponding to the ground-state energy of the spin system $g(T=0,p;\delta)=\alpha\delta^2/2+p\delta+e_0(\delta)$,
and we may argue 
about the effect of the spin system on the observed lattice compressibility $1/\varkappa$ defined in Eq. (\ref{2.03}).
It is worthwhile to consider briefly several well-known spin models.

{\it {Ising chain in a longitudinal field\cite{arbitrary}.}}
We begin with the spin Hamiltonian
\begin{eqnarray}
\label{2.04}
H=\sum_{n}\left(Js_n^{z}s_{n+1}^{z}-hs_n^{z}\right),
\;\;\;
h\ge 0
\end{eqnarray}
and assume $J=J_0(1-k\delta)$ with $k=1$ for simplicity.
The adopted linear dependence of $J$ on $\delta$ can be justified only if $\delta\ll 1$.

For the ferromagnetic exchange interaction $J_0=-\vert J_0\vert <0$
the ground-state energy per site is given by $e_0(\delta)=-\vert J_0\vert(1-\delta)/4 -h/2$
that results in a new equilibrium value of $\delta$, $\delta(p)=-(p+\vert J_0\vert/4)/\alpha$,
but leaves the inverse compressibility unchanged, $\varkappa=\alpha$.
\{Note that for $p=0$ we have $\delta(0)=-\vert J_0\vert/(4\alpha)$, 
i.e., the lattice shrinks and the ferromagnetic exchange constant $J=J_0[1+\vert J_0\vert/(4\alpha)]$ becomes larger than $J_0$.\}

For the antiferromagnetic exchange interaction $J_0=\vert J_0\vert >0$, 
the ground-state energy per site is given 
either by 
$e_{0{\rm{w}}}(\delta )=-\vert J_0\vert (1-\delta)/4$ if $h<\vert J_0\vert(1-\delta)$
(weak-field regime) 
or by
$e_{0{\rm{s}}}(\delta)=\vert J_0\vert (1-\delta)/4-h/2$ if $ h > \vert J_0\vert(1-\delta)$
(strong-field regime). 
Let us introduce the critical fields 
$h_c/\vert J_0\vert=1+p/\alpha$, $h_{c1}/\vert J_0\vert=h_c/\vert J_0\vert-\vert J_0\vert/(4\alpha)$,
and 
$h_{c2}/\vert J_0\vert=h_c/\vert J_0\vert+\vert J_0\vert/(4\alpha)$,
the critical pressures 
$p_c=(h/\vert J_0\vert-1)\alpha$, $p_{c1}=p_c-\vert J_0\vert/4$, 
and $p_{c2}=p_c+\vert J_0\vert/4$,
and the two equilibrium values of 
$\delta$, $\delta_1(p)=-(p+\vert J_0\vert/4)/\alpha$
and 
$\delta_2(p)=-(p-\vert J_0\vert/4)/\alpha$.

Considering only the global minimum of the variational enthalpy $g(T=0,p,h;\delta)$ 
one arrives at the following conclusions. 
{\it{i)}}
At fixed values of $p$, the equilibrium value of $\delta$ is 
either $\delta_1(p)$ for $h < h_c$ or $\delta_2(p)$ for $h > h_c$ 
and there is a jump in between these two equilibrium values at the critical field $h=h_c$.
{\it{ii)}}
Under the constant field $h$, the equilibrium value of $\delta$ is 
$\delta_2(p)$ for $p < p_c$, $\delta_1(p)$ for $p > p_{c}$ 
and the abrupt change in $\delta(p)$ appears at the critical pressure $p=p_c$. 

However, it should be also stressed that the dependence of variational enthalpy $g(T=0,p,h;\delta)$ on $\delta$ 
may exhibit an additional local minimum besides the global one 
when driving the antiferromagnetic Ising chain sufficiently close to the critical field $h_c$ or the critical pressure $p_c$. 
This implies a possibility of observing hysteresis phenomena 
originating from the first-order phase transitions driven either by varying of field $h$ or pressure $p$. 
Considering the metastable states, which correspond to the local minima of the variational enthalpy, 
one arrives after a simple analysis at the following conclusions.
{\it{i)}}
For fixed $p$, 
the equilibrium value of $\delta$ is $\delta_1(p)$ when $h$ increases from 0 to $h_{c2}$
and then it jumps to $\delta_2(p)$ when $h$ further increases
(increasing-field regime).
However, the equilibrium value of $\delta$ is $\delta_2(p)$ when $h$ decreases from $\infty$ to $h_{c1}$
and then it jumps to $\delta_1(p)$ upon further decrease of $h$
(decreasing-field regime).
Furthermore, the relations $\partial\delta_1(p)/\partial p=\partial\delta_2(p)/\partial p=-1/\alpha$ give $\varkappa=\alpha$.
{\it{ii)}}
For fixed $h$,
the equilibrium value of $\delta$ is $\delta_2(p)$ when $p$ varies from $-\infty$ to $p_{2c}$ 
and then it jumps to $\delta_1(p)$ as $p$ exceeds $p_{2c}$ (increasing-pressure regime).
On the other hand,
the equilibrium value of $\delta$ is $\delta_1(p)$ when $p$ varies from $\infty$ to $p_{1c}$ 
and then it jumps to $\delta_2(p)$ whenever $p$ becomes less than $p_{1c}$ (decreasing-pressure regime).
Furthermore, $\varkappa=\alpha$ unless $p\ne p_{c2}$ but $\varkappa(p_{c2})=0$ for the increasing-pressure regime,
whereas $\varkappa=\alpha$ unless $p\ne p_{c1}$ but $\varkappa(p_{c1})=0$ for the decreasing-pressure regime.

{\it {$XX$ chain in a transverse field.}}
We pass to the one-dimensional spin-1/2 system with isotropic $XY$ interaction in the presence of a transverse magnetic field.
The Hamiltonian of the spin model reads
\begin{eqnarray}
\label{2.05}
H=\sum_{n}\left[J\left(s_n^{x}s_{n+1}^{x}+s_n^{y}s_{n+1}^{y}\right)-hs_n^{z}\right],
\;\;\;
h\ge 0
\end{eqnarray}
and $J=J_0(1-\delta)$.
The sign of $J$ is not important for the thermodynamic quantities and therefore, we may set $J_0>0$ without loss of generality.
Although this model in the present context was discussed in some detail in Ref.~\onlinecite{orignac},
we report here a brief account of to some extent known results 
for self-consistent readability of our paper and for reader's convenience.
The magnetic contribution $e_0(\delta)$ to the variational enthalpy 
can easily be found using the Jordan-Wigner fermionization approach\cite{lieb,katsura}.
Separating the weak-field regime and the strong-field regime we have
\begin{eqnarray}
\label{2.06}
e_0(\delta) &=&
\left\{
\begin{array}{ll}
e_{0{\rm{w}}}(\delta), & h < J_0(1-\delta),\\
e_{0{\rm{s}}}(\delta), & h > J_0(1-\delta),
\end{array}
\right.
\nonumber\\
e_{0{\rm{w}}}(\delta)&=&-\frac{J_0(1-\delta)}{\pi}\sin k_{\rm F} - h\left(\frac{1}{2}-\frac{k_{\rm F}}{\pi}\right),
\nonumber\\
e_{0{\rm{s}}}(\delta)&=&-\frac{h}{2}
\end{eqnarray}
with $\cos k_{\rm F}=h/[J_0(1-\delta)]$. 
Again, $g(T=0,p,h;\delta)$ versus $\delta$ dependence can have two minima 
suggesting a possible existence of first-order phase transitions accompanied by hysteresis phenomena.

More specifically, consider the equation for $\delta(p)$ in the weak-field regime 
\begin{eqnarray}
\label{2.07}
\alpha\delta+p+\frac{J_0}{\pi}\sqrt{1-\frac{h^2}{J_0^2(1-\delta)^2}}=0.
\end{eqnarray}
Fixing $p$ and varying $h$ from 0 to $\infty$ one finds that 
{\it i)} 
Eq. (\ref{2.07}) has only one solution 
which corresponds to a minimum of $g_{{\rm{w}}}(T=0,p,h;\delta)=\alpha\delta^2/2+p\delta+e_{0{\rm{w}}}(\delta)$
for any $h<h_{c1}=(1+p/\alpha)J_0$;
{\it ii)} 
at $h=h_{c1}$ a second solution (at $-p/\alpha$) 
corresponding to a maximum of $g_{{\rm{w}}}(T=0,p,h;\delta)$ emerges; 
{\it iii)}
Eq. (\ref{2.07}) has two solutions for $h_{c1}<h<h_{c2}$ 
that are gradually approaching each other as $h$ increases;
{\it iv)}
at $h=h_{c2}$ two solutions coincide and Eq. (\ref{2.07}) has no appropriate solutions for larger $h$.
Let us denote the smaller root of Eq. (\ref{2.07}) by $\delta_1(p)$.
The equation for $\delta(p)$ in the strong-field regime yields $\delta_2(p)=-p/\alpha$
which corresponds to a minimum of $g_{{\rm{s}}}(T=0,p,h;\delta)=\alpha\delta^2/2+p\delta+e_{0{\rm{s}}}(\delta)$.
One more characteristic field, $h_{c1}<h_c<h_{c2}$, is determined from the condition
$g_{{\rm{w}}}(T=0,p,h_c;\delta_1(p))=g_{{\rm{s}}}(T=0,p,h_c;\delta_2(p))$.

Now one may immediately conclude that the equilibrium value of $\delta$ 
is $\delta_1(p)$ for $h<h_{c}$ and $\delta_2(p)$ for $h>h_{c}$ 
when considering the thermodynamically stable solution corresponding to the global minimum of the variational enthalpy only. 
On the other hand, 
the inclusion of metastable states into our analysis allows one to study an interesting hysteresis phenomena 
resulting from the magneto-elastic coupling.
Under this circumstance, 
one actually finds that the equilibrium value of $\delta$ is $\delta_1(p)$ when $h$ increases from 0 to $h_{c2}$ 
and then it jumps to $\delta_2(p)>\delta_1(p)$ when $h$ further increases 
(increasing-field regime).
However, the equilibrium value of $\delta$ is $\delta_2(p)$ when $h$ decreases from $\infty$ to $h_{c1}$
and then it jumps to $\delta_1(p)<\delta_2(p)$ upon further decrease of $h$ 
(decreasing-field regime).
The inverse compressibility for the increasing-field regime follows
\begin{eqnarray}
\label{2.08}
\varkappa=\alpha-\frac{h^2}{\pi\left(1-\delta_1(p)\right)^2\sqrt{J_0^2(1-\delta_1(p))^2-h^2}}
\nonumber\\
=\alpha+\frac{h^2}{\pi^2\left(\alpha\delta_1(p)+p\right)\left(1-\delta_1(p)\right)^3}
\end{eqnarray}
until $h<h_{c2}$, but $\varkappa=\alpha$ if $h>h_{c2}$. 
For the decreasing-field regime $\varkappa=\alpha$ until $h>h_{c1}$ but $\varkappa$ follows Eq. (\ref{2.08}) if $h<h_{c1}$.

Similarly, fixing $h$ and varying $p$ from $-\infty$ to $\infty$ one finds that 
{\it i)} 
Eq. (\ref{2.07}) has no appropriate solutions for $p<p_{c1}$;
{\it ii)} 
Eq. (\ref{2.07}) has two solutions if $p$ exceeds $p_{c1}$ 
and the smaller one, $\delta_1(p)$, corresponds to a minimum of $g_{{\rm{w}}}(T=0,p,h;\delta)$;
{\it iii)} 
the larger solution (at $-p/\alpha$) corresponding to a maximum of $g_{{\rm{w}}}(T=0,p,h;\delta)$ disappears at $p=p_{c2}$; 
{\it iv)} Eq. (\ref{2.07}) has only one solution $\delta_1(p)$ for $p>p_{c2}$.
One more characteristic pressure, $p_{c1}<p_c<p_{c2}$, is determined from the condition
$g_{{\rm{w}}}(T=0,p_c,h;\delta_1(p_c))=g_{{\rm{s}}}(T=0,p_c,h;\delta_2(p_c))$.

Considering only the stable solution with the lowest variational enthalpy 
one finds that the equilibrium value of $\delta$ is either $\delta_2(p)$ if $p<p_c$ or $\delta_1(p)$ if $p>p_c$.
It is also quite obvious from the aforementioned analysis 
that the striking pressure-induced hysteresis phenomenon emerges when taking into consideration an existence of the metastable states.
Indeed, it can be readily proved 
that the equilibrium value of $\delta$ is $\delta_2(p)$ when $p$ increases from 0 to $p_{c2}$ 
and then it jumps to $\delta_1(p)<\delta_2(p)$ when $p$ further increases (increasing-pressure regime).
On the other hand, 
the equilibrium value of $\delta$ is given by $\delta_1(p)$ when $p$ decreases from $\infty$ to $p_{c1}$
and then it jumps to $\delta_2(p)>\delta_1(p)$ upon further decrease of $p$ (decreasing-pressure regime).
Furthermore,
$\varkappa=\alpha$ for the increasing-pressure regime until $p<p_{c2}$ but $\varkappa$ follows Eq. (\ref{2.08}) if $p>p_{c2}$.
For the decreasing-field regime the inverse compressibility follows Eq. (\ref{2.08}) if $p>p_{c1}$ but $\varkappa=\alpha$ if $p<p_{c1}$.

\begin{figure}[htb]
\begin{center}
\includegraphics[width=9cm]{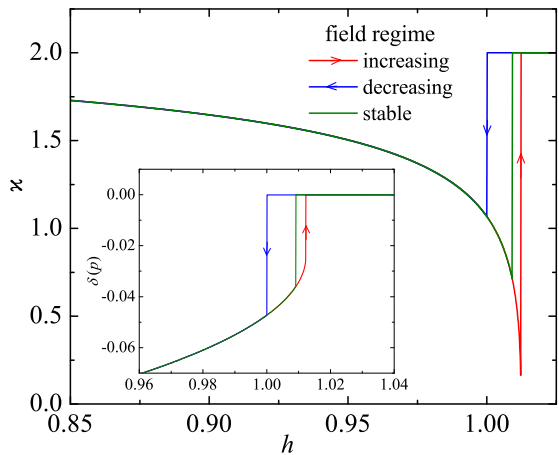}
\end{center}
\vspace{-1.2cm}
\caption{(Color online). 
Inverse compressibility $\varkappa$ versus field $h$ 
for the spin-1/2 $XX$ chain in a transverse field (\ref{2.05}) with $J_0=1$, $\alpha=2$ at $p=0$. 
The line with up-pointing arrow (in red) corresponds to the increasing-field regime, 
the line with down-pointing arrow (in blue) corresponds to the decreasing-field regime, 
and the line without an arrow (in green) corresponds to the global minimum of the variational enthalpy.
Inset: Corresponding $\delta(p=0)$ versus $h$ dependences.}
\label{fig_01}
\end{figure}
\begin{figure}[htb]
\begin{center}
\includegraphics[width=9cm]{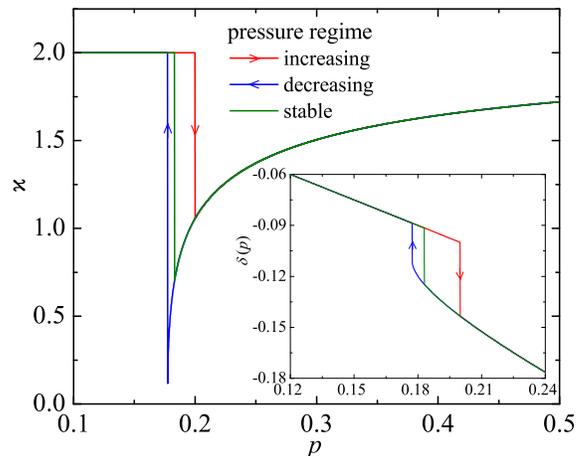}
\end{center}
\vspace{-1.2cm}
\caption{(Color online). 
Inverse compressibility $\varkappa$ versus pressure $p$ 
for the spin-1/2 $XX$ chain in a transverse field (\ref{2.05}) with $J_0=1$, $\alpha=2$ at $h=1.1$. 
The line with down-pointing arrow (in red) corresponds to the increasing-pressure regime, 
the line with up-pointing arrow (in blue) corresponds to the decreasing-pressure regime, 
and the line without an arrow (in green) corresponds to the global minimum of the variational enthalpy.
Inset: Corresponding $\delta(p)$ versus $p$ dependences.}
\label{fig_02}
\end{figure}

In Figs.~\ref{fig_01} and \ref{fig_02} we demonstrate dependences of the inverse compressibility $\varkappa$ on $h$ and $p$
along with corresponding dependences of the equilibrium value of $\delta$ depicted in the insets. 
It is noteworthy that the displayed figures may serve in evidence of the aforedescribed field- and pressure-induced hysteresis phenomena 
and besides, one also observes here an abrupt jump of $\varkappa$ and $\delta(p)$ at the first-order phase transitions. 
The interested reader is referred to the paper by Orignac and Citro\cite{orignac} 
for a more complete discussion of the deformable spin-1/2 $XX$ chain in a transverse field.

A brief summary of this section is as follows.
Although the equilibrium value of $\delta(p)$ may vary after placing interacting spins on a lattice,
the observed elastic constant $\varkappa$ does not necessarily feel the appearance of interacting spins
as for instance in the spin-1/2 ferromagnetic Ising chain in a longitudinal field. 
Contrary to this, the spin-1/2 $XX$ chain in a transverse field 
clearly shows a ground-state peculiarity of the quantum spin model in the elastic properties of the underlying lattice:
the inverse compressibility $\varkappa$ noticeably diminishes 
while the spin chain approaches a quantum critical point indicating a change 
from the spin-liquid phase to the ferromagnetic phase.
More precisely,
the inverse compressibility $\varkappa$ shows a jump to a finite value accompanied with hysteresis.
Formally, a specific behavior of the elastic properties arises from a specific dependence $e_0(\delta)$ for the spin model at hand at a quantum critical point.

\section{Ising chain in a transverse field}
\label{sec3}
\setcounter{equation}{0}

In this section, 
we consider the one-dimensional spin-1/2 transverse Ising chain given by the Hamiltonian
\begin{eqnarray}
\label{3.01}
H=\sum_{n}\left(Js_n^{x}s_{n+1}^{x}-hs_n^{z}\right),
\;\;\;
h\ge 0
\end{eqnarray}
with $J=J_0(1-\delta)$.
Again we may assume $J_0>0$ without loss of generality.
This model, 
although to our knowledge has not been considered so far,
can be again studied rigorously using the Jordan-Wigner fermionization\cite{lieb,katsura}.
The magnetic ground-state energy per site reads
\begin{eqnarray}
\label{3.02}
e_0(\delta)=-\frac{J_0(1-\delta)+2h}{2\pi}{\bf{E}}(z),
\nonumber\\
z^2=1-\left[\frac{J_0(1-\delta)-2h}{J_0(1-\delta)+2h}\right]^2,
\end{eqnarray}
where
${\bf{E}}(z)\equiv\int_0^{\pi/2}d\phi\sqrt{1-z^2\sin^2\phi}$
is the complete elliptic integral of the second kind with the modulus $z$ defined through Eq. (\ref{3.02}). 
Recall\cite{jahnke_emde_loesch,oldham} that around $z\approx 1$ we have
${\bf{E}}(z)\approx 1+(1/2)[\ln(4/\sqrt{1-z^2})-1/2](1-z^2)$,
that implies possible peculiarities when $J_0(1-\delta)-2h=0$.
It can be readily verified 
that the dependence of variational enthalpy $g(T=0,p,h;\delta)=\alpha\delta^2/2+p\delta+e_0(\delta)$ on the distortion parameter $\delta$ 
has only one minimum at $\delta(p)$, 
whereas the value of $\delta(p)$ obeys the condition
\begin{eqnarray}
\label{3.03}
\alpha\delta+p
+\frac{\partial e_0(\delta)}{\partial\delta}=0,
\end{eqnarray}
and the inverse compressibility then follows from
\begin{eqnarray}
\label{3.04}
\varkappa=\alpha+\left.\frac{\partial^2e_0(\delta)}{\partial\delta^2}\right\vert_{\delta=\delta(p)}.
\end{eqnarray}
Here, $e_0(\delta)$ is given by Eq. (\ref{3.02}).

\begin{figure}[htb]
\begin{center}
\includegraphics[width=9cm]{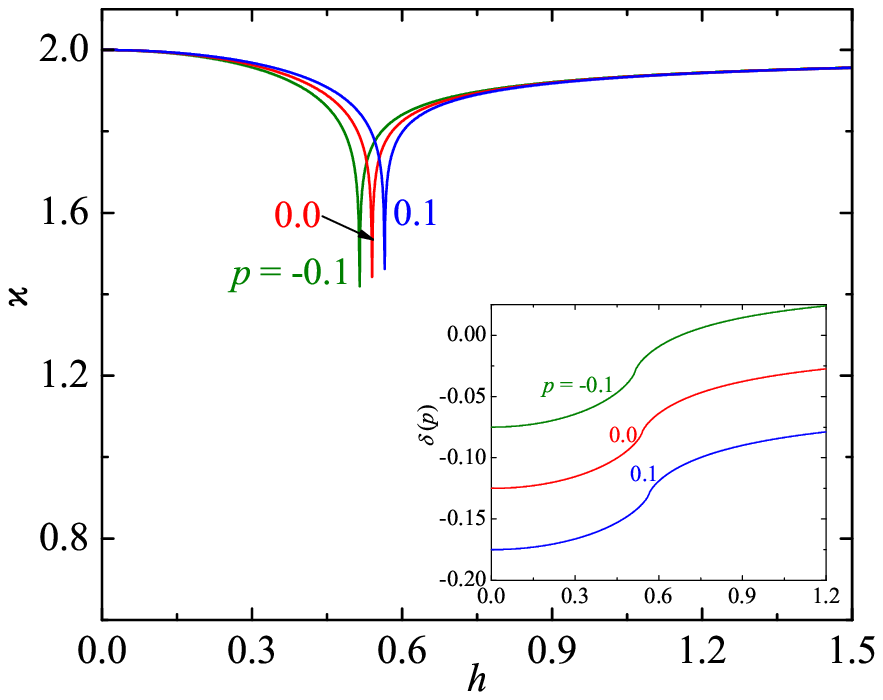}
\end{center}
\vspace{-1.2cm}
\caption{(Color online). 
Inverse compressibility $\varkappa$ versus field $h$ for the spin-1/2 Ising chain in a transverse field (\ref{3.01}) 
with $J_0=1$, $\alpha=2$ at $p=-0.1$ (green line), $p=0$ (red line), and $p=0.1$ (blue line).
Inset: Corresponding $\delta(p)$ versus $h$ dependences.}
\label{fig_03}
\end{figure}
\begin{figure}[htb]
\begin{center}
\includegraphics[width=9cm]{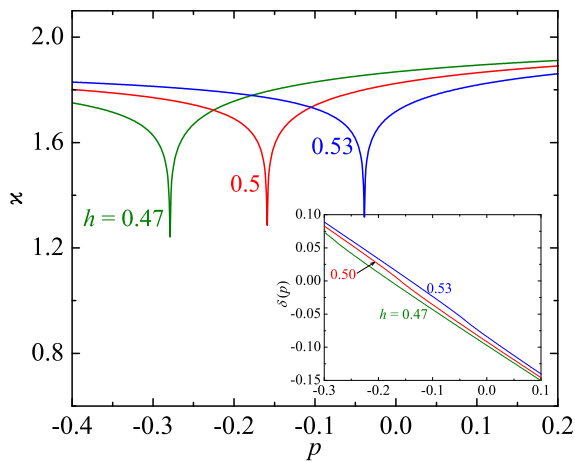}
\end{center}
\vspace{-1.2cm}
\caption{(Color online). 
Inverse compressibility $\varkappa$ versus pressure $p$ for the spin-1/2 Ising chain in a transverse field (\ref{3.01}) 
with $J_0=1$, $\alpha=2$ at $h=0.47$ (green line), $h=0.5$ (red line), and $h=0.53$ (blue line).
Inset: Corresponding $\delta(p)$ versus $p$ dependences.}
\label{fig_04}
\end{figure}

In Figs.~\ref{fig_03} and \ref{fig_04} we demonstrate dependences $\varkappa(h)$ at fixed $p$ and $\varkappa(p)$ at fixed $h$ 
along with corresponding dependences of the equilibrium value of $\delta$
as they follow from the numerical solution of Eqs. (\ref{3.03}) and (\ref{3.04}) with $e_0(\delta)$ (\ref{3.02}),
which have been solved by subsequent numerical integration and differentiation.
Clearly, if the minimum of $g(T=0,p,h;\delta)$, which occurs at $\delta(p)$ given by Eq. (\ref{3.03}),
satisfies the relation $J_0(1-\delta(p))-2h=0$,
i.e., when $\delta(p)=\delta_c$ with $\delta_c=1-2h/J_0$,
peculiarities in various quantities may be expected.
For fixed $p$ and varying $h$ we find the critical field $h_c/J_0=1/2+[p+J_0/(2\pi)]/(2\alpha)$
at which $\delta(p)=\delta_c=1-2h_c/J_0=-[p+J_0/(2\pi)]/\alpha$.
For fixed $h$ and varying $p$ we find the critical pressure $p_c=-\alpha(1-2h/J_0)-J_0/(2\pi)$ 
at which $\delta(p_c)=\delta_c=1-2h/J_0$.
It is noteworthy 
that the most obvious changes in the monotonous $\delta(p)$ dependence on $h$ (on $p$) at fixed $p$ (at fixed $h$) 
appear close to the critical field $h_c$ (critical pressure $p_c$),
see the insets of Figs.~\ref{fig_03} and \ref{fig_04}. 

Let us take a closer look on Eqs. (\ref{3.02}) -- (\ref{3.04}).
For this purpose we rewrite Eq. (\ref{3.03}) in the form
\begin{eqnarray}
\label{3.05}
\alpha \delta + p 
\nonumber\\
+\frac{J_0(1-\delta)+2h}{4\pi(1-\delta)}{\bf{E}}(z)
+\frac{J_0(1-\delta)-2h}{4\pi(1-\delta)}{\bf{K}}(z)
=0, 
\end{eqnarray}
where ${\bf{K}}(z)\equiv\int_0^{\pi/2}d\phi/\sqrt{1-z^2\sin^2\phi}$
is the complete elliptic integral of the first kind
which behaves as ${\bf{K}}(z)\approx\ln(4/\sqrt{1-z^2})$ around $z\approx 1$\cite{jahnke_emde_loesch,oldham},
and Eq. (\ref{3.04}) in the form
\begin{eqnarray}
\label{3.06}
\varkappa=\alpha 
+\frac{h}{2\pi(1-\delta)^2}\left[{\bf{E}}(z)-{\bf{K}}(z)\right]
\nonumber\\
+\left[\frac{J_0(1-\delta)+2h}{4\pi(1-\delta)}\frac{d{\bf{E}}(z)}{dz}
+\frac{J_0(1-\delta)-2h}{4\pi(1-\delta)}\frac{d{\bf{K}}(z)}{dz}\right]\frac{dz}{d\delta}.
\end{eqnarray}
Here, $z$ is given by Eq. (\ref{3.02}),
$d{\bf{E}}(z)/dz=[{\bf{E}}(z)-{\bf{K}}(z)]/z$,
$d{\bf{K}}(z)/dz=-{\bf{K}}(z)/z+{\bf{E}}(z)/[z(1-z^2)]$,
and $dz/d\delta=4J_0h[J_0(1-\delta)-2h]/\{z[J_0(1-\delta)+2h]^3\}$.
Clearly,
Eqs. (\ref{3.05}) and (\ref{3.06}) immediately yield the dependences $p(\delta)$ and $\varkappa(\delta)$, respectively.

Consider now, for instance, the case when $h$ is fixed and $p$ varies.
Inverting the dependence $p=p(\delta)$ which follows from Eq. (\ref{3.05}) 
we may obtain $\delta(p)$ (see the inset in Fig.~\ref{fig_04}),
whereas combining the dependences $p=p(\delta)$ and $\varkappa=\varkappa(\delta)$ which follow from Eqs. (\ref{3.05}) and (\ref{3.06})
we may obtain $\varkappa(p)$ (see the main panel in Fig.~\ref{fig_04}).

If $\delta=\delta_c+\varepsilon$ is close to $\delta_c$
(i.e., $\vert\varepsilon\vert\to 0$),
$\sqrt{1-z^2}\approx \vert\varepsilon\vert J_0/(4h)$ 
and Eq. (\ref{3.05}) becomes
\begin{eqnarray}
\label{3.07}
p-p_c\approx\left(-\alpha+J_0^2\frac{\ln\frac{16h}{J_0}-1}{8\pi h}
-\frac{J_0^2}{8\pi h}\ln\vert\varepsilon\vert\right)\varepsilon
\nonumber\\
=-{\cal{C}}\frac{J_0^2}{8\pi h}\epsilon\ln\vert\epsilon\vert,
\nonumber\\
\ln{\cal{C}}=-\frac{8\pi\alpha h}{J_0^2}+\ln\frac{16h}{J_0}-1,
\;\;\;
\epsilon=\frac{\varepsilon}{{\cal{C}}}.
\end{eqnarray}
Although $\vert\varepsilon\vert$ in Eq. (\ref{3.07}) is small,
it is not less than ${\cal{C}}$ 
(for $h/J_0=0.45\ldots 0.55$, $\alpha/J_0=2$ we have ${\cal{C}}\approx 3.98\cdot 10^{-10}\ldots 3.19\cdot 10^{-12}$),
i.e., $\vert\epsilon\vert\ge 1$,
in order not to violate the relation ${\rm{sgn}}(p-p_c)=-{\rm{sgn}}(\varepsilon)$.
Differentiating Eq. (\ref{3.07}) with respect to $\delta$ we obtain
$\varkappa=[J_0^2/(8\pi h)](\ln\vert\epsilon\vert+1)$
and $\varkappa$ falls down to $J_0^2/(8\pi h)$ at $\vert\epsilon\vert=1$.
However, these values of $\varkappa$ cannot be reached numerically, cf. Fig.~\ref{fig_04}. 

A brief summary of this section is as follows.
Both the equilibrium value of $\delta(p)$ and the observed elastic constant $\varkappa$
clearly show a ground-state peculiarity of the quantum spin model put on the lattice.
In particular,
the inverse compressibility $\varkappa$ is noticeably diminished 
while the spin chain approaches a quantum critical point indicating a change from the Ising phase to the paramagnetic phase.
In contrast to the transverse $XX$ chain,
the inverse compressibility $\varkappa$ for the transverse Ising chain decreases continuously but not discontinuously.

Now we apply the elaborated scheme
to the spin-1/2 Heisenberg-Ising bond alternating chain refined by an additional Dzyaloshinskii-Moriya interaction.
The model exhibits a quantum phase transition point driven by a relation between the interaction constants,
which separates the disordered phase (weak Ising interaction) and the long-range ordered phase (strong Ising interaction),
see Ref.~\onlinecite{lieb}.
Our task is to follow how this quantum critical point manifests itself in the elastic properties.

\section{Heisenberg-Ising chain with Dzyaloshinskii-Moriya interaction}
\label{sec4}
\setcounter{equation}{0}

In this section, we focus on the ground-state properties
of a regularly alternating spin-1/2 antiferromagnetic Heisenberg-Ising chain with a Dzyaloshinskii-Moriya interaction.
The chain consists of $N=2{\cal{N}}$ sites with spins $s=1/2$.
Moreover,
the $XXZ$ Heisenberg interaction and the $z$-component of the Dzyaloshinskii-Moriya interaction
regularly interchange with the Ising interaction in this chain.
The Hamiltonian of the spin model reads
\begin{eqnarray}
\label{4.01}
H = \sum_{m=1}^{{\cal{N}}}\left(H_{2m-1,2m}+H_{2m,2m+1}\right)
\end{eqnarray}
with
\begin{eqnarray}
\label{4.02}
H_{2m-1,2m}
&=&J_{\rm H}\left(s^x_{2m-1}s^x_{2m}+s^y_{2m-1}s^y_{2m}+\Delta s^z_{2m-1}s^z_{2m}\right)
\nonumber\\
&+&D\left(s^x_{2m-1} s^y_{2m}-s^y_{2m-1} s^x_{2m}\right)
\end{eqnarray}
and
\begin{eqnarray}
\label{4.03}
H_{2m,2m+1}=J_{\rm I}s_{2m}^z s_{2m+1}^z.
\end{eqnarray}
Here the parameters $J_{\rm H}>0$ and $D$ denote the antiferromagnetic Heisenberg and the Dzyaloshinskii-Moriya interactions 
between $2m-1$ and $2m$ spins,
$\Delta$ controls the anisotropy of the Heisenberg interaction,
and the parameter $J_{\rm I}>0$ denotes the antiferromagnetic Ising interaction between $2m$ and $2m+1$ spins.
We impose periodic boundary conditions for convenience.

Although a merit of the introduced quantum spin chain is possibility to perform accurate calculations,
it might be worthy to notice here that mixed Heisenberg-Ising interactions do occur 
in some magnetic compounds containing lanthanide ions\cite{dysprosium}.
Moreover, the magnetic subsystem of such compounds may have one-dimensional geometry.

The Hamiltonian (\ref{4.01}), (\ref{4.02}), (\ref{4.03}) corresponds to the exactly solvable Heisenberg-Ising model
proposed in Ref.~\onlinecite{lieb} (see also Refs.~\onlinecite{lieh,yao,csmag,taras-jozef}).
The calculation of the ground-state energy of the model in brief looks as follows.
First we eliminate the Dzyaloshinskii-Moriya term from the Hamiltonian
after performing an appropriate spin-coordinate transformation\cite{dm1,dm2,dm3},
\begin{eqnarray}
\label{4.04}
s_{2m}^x\to  s_{2m}^x\cos\varphi+s_{2m}^y\sin\varphi,
\nonumber\\
s_{2m}^y\to -s_{2m}^x\sin\varphi+s_{2m}^y\cos\varphi
\end{eqnarray}
with $\tan\varphi=D/J_{\rm{H}}$
for $m=1,\ldots,{\cal{N}}$,
which results in
\begin{eqnarray}
\label{4.05}
H_{2m-1,2m}
&=& J_{XY}\left(s^x_{2m-1}s^x_{2m}+s^y_{2m-1}s^y_{2m}\right)
\nonumber\\
&+& J_{\rm H}\Delta s^z_{2m-1}s^z_{2m},
\nonumber\\
J_{XY} &=& \sqrt{J^2_{\rm H}+D^2}.
\end{eqnarray}
From Eq. (\ref{4.05}) we see that the introduced Dzyaloshinskii-Moriya interaction
effectively increases the $XY$ component of the Heisenberg interaction with respect to the Ising component of this interaction.

Then we follow the arguments of Refs.~\onlinecite{lieb,taras-jozef}.
The ground state of the antiferromagnetic model (\ref{4.01}), (\ref{4.05}), (\ref{4.03}) must lie in a particular subspace, 
where all Heisenberg bonds are in one of two states:
either
$(\vert\downarrow\rangle_{2m-1}\vert\uparrow\rangle_{2m}
+\vert\uparrow\rangle_{2m-1}\vert\downarrow\rangle_{2m})/\sqrt{2}$
or
$(\vert\downarrow\rangle_{2m-1}\vert\uparrow\rangle_{2m}
-\vert\uparrow\rangle_{2m-1}\vert\downarrow\rangle_{2m})/\sqrt{2}$.
Introducing the raising and lowering operators for the $m$th pair, 
$a_m^{\dagger}$ and $a_m$, $m=1,\ldots,{\cal{N}}$,
which satisfy the Fermi commutation relations if they are attached to the same site,
but obviously commute if they are attached to different sites,
the Hamiltonian (\ref{4.01}), (\ref{4.05}), (\ref{4.03}) becomes
\begin{eqnarray}
\label{4.06}
H=&-&\frac{{\cal{N}}}{2}J_{XY}-\frac{{\cal{N}}}{4}J_H\Delta 
+\sum_{m=1}^{\cal{N}}\Biggl[J_{XY}a_m^{\dagger}a_m
\nonumber\\
&-&\frac{J_{\rm{I}}}{4}\left(a_m^{\dagger}+a_m\right)\left(a_{m+1}^{\dagger}+a_{m+1}\right)\Biggr].
\end{eqnarray}
After applying the Jordan-Wigner transformation\cite{jordan},
\begin{eqnarray}
\label{4.07}
c_1=a_1,
\;\;\;
c_m={\rm e}^{i\pi\sum_{l=1}^{m-1}a_l^{\dagger}a_l}a_m,
\;
m=2,\ldots,{\cal{N}},
\end{eqnarray}
Eq. (\ref{4.06}) becomes a bilinear form in Fermi operators
\begin{eqnarray}
H=&-&\frac{{\cal{N}}}{2}J_{XY}-\frac{{\cal{N}}}{4}J_H\Delta +\sum_{m=1}^{\cal{N}}\Biggl[J_{XY}c_m^{\dagger}c_m
\nonumber\\ 
\label{4.08}
&-&\frac{J_{\rm{I}}}{4}\left(c_m^{\dagger}-c_m\right)\left(c_{m+1}^{\dagger}+c_{m+1}\right)\Biggr].
\end{eqnarray}
We assume periodic boundary conditions in Eq. (\ref{4.08})
omitting an unimportant boundary term in the limit ${\cal{N}}\to\infty$ for further calculations.
To bring the Hamiltonian to a diagonal form we perform first the Fourier transformation,
\begin{eqnarray}
\label{4.09}
c_{\kappa}=\frac{1}{\sqrt{{\cal{N}}}}\sum_{m=1}^{{\cal{N}}} {\rm e}^{i\kappa m}c_m,
\end{eqnarray}
$\kappa=2\pi l/{\cal{N}}$,
$l=-{\cal{N}}/2,\ldots,{\cal{N}}/2-1$
(we assume ${\cal{N}}$ is even),
and then the Bogolyubov transformation,
\begin{eqnarray}
\label{4.10}
\beta_{\kappa}=i\sin g_{\kappa} c_{\kappa}+\cos g_{\kappa} c^{\dagger}_{-\kappa},
\nonumber\\
\tan(2g_{\kappa})=\frac{J_{\rm{I}}\sin\kappa}{2J_{XY}-J_{\rm{I}}\cos\kappa}.
\end{eqnarray}
The Hamiltonian (\ref{4.08}) becomes
\begin{eqnarray}
\label{4.11}
H&=&-\frac{{\cal{N}}}{4}J_{\rm{H}}\Delta+\sum_{\kappa}\Lambda_{\kappa}
\left(\beta_{\kappa}^{\dagger}\beta_{\kappa}-\frac{1}{2}\right),
\nonumber\\
\Lambda_{\kappa}
&=& \sqrt{\left(J_{XY}+\frac{1}{2}J_{\rm{I}}\right)^2-2J_{XY}J_{\rm{I}}\cos^2\frac{\kappa}{2}}.
\end{eqnarray}
Since Eq. (\ref{4.11}) is the Hamiltonian which acts in the subspace
to which the ground state of the spin model (\ref{4.01}), (\ref{4.05}), (\ref{4.03}) belongs
we immediately find the required ground-state energy per site in the thermodynamic limit
\begin{eqnarray}
\label{4.12}
e_0&=&-\frac{1}{8}J_{\rm{H}}\Delta
-\frac{J_{XY}+\frac{1}{2}J_{\rm{I}}}{2\pi}{\bf{E}}(z),
\nonumber\\
z^2&=&\frac{2J_{XY}J_{\rm{I}}}{\left(J_{XY}+\frac{1}{2}J_{\rm{I}}\right)^2}.
\end{eqnarray}
Here ${\bf{E}}(z)$ is the complete elliptic integral of the second kind (see Sec.~\ref{sec3}) with the modulus $z$ given by (\ref{4.12}).

It is worth to recall some ground-state properties of the spin model 
defined through the Hamiltonians (\ref{4.01}), (\ref{4.02}), (\ref{4.03}). 
The energy spectrum $\Lambda_{\kappa}$ is gapped unless
\begin{eqnarray}
\label{4.13}
J_{\rm{I}}=2J_{XY}=2\sqrt{J_{\rm{H}}^2+D^2}.
\end{eqnarray}
If the condition (\ref{4.13}) holds the spin model has a gapless excitation spectrum\cite{csmag}.
From Eq. (\ref{4.13}) we see that only the relation between the Ising interaction $J_{\rm{I}}$ 
and the effective $XY$ component of the Heisenberg interaction $J_{XY}$ is relevant for adjusting to a critical point, 
whereas the latter effective interaction $J_{XY}$ can be altered by varying the Dzyaloshinskii-Moriya interaction $D$.
Furthermore, the condition (\ref{4.13}) yields $z^2=1$ in Eq. (\ref{4.12})
thus implying peculiarities in the ground-state properties of the investigated bond alternating chain.

Now we turn to a deformable spin-1/2 Heisenberg-Ising chain with Dzyaloshinskii-Moriya interaction
and assume that the interspin interactions depend on the change of intersite distance as follows:
$J_{\rm{H}}={\cal{J}}_{\rm{H}}(1-k_{\rm{H}}\delta)$,
$D={\cal{D}}(1-k_{\rm{D}}\delta)$,
$J_{\rm{I}}={\cal{J}}_{\rm{I}}(1-k_{\rm{I}}\delta)$.
We introduce the parameters $k_{\rm{H}}$, $k_{\rm{D}}$ and $k_{\rm{I}}$
to distinguish between the effects of the distance change on different intersite interactions.
Using Eq. (\ref{4.12}) we may obtain the magnetic contribution $e_0(\delta)$ to the variational enthalpy per site $g(T=0,p;\delta)$
\begin{widetext}
\begin{eqnarray}
\label{4.14}
\!\!\!\!&&\!\!\!\! e_0(\delta)=
-\frac{1}{8}{\cal{J}}_{\rm{H}}\Delta(1-k_{\rm{H}}\delta)
\nonumber\\
\!\!\!\!&-&\!\!\!\! \frac{2 \sqrt{{\cal{J}}^2_{\rm{H}}(1-k_{\rm{H}}\delta)^2+{\cal{D}}^2(1-k_{\rm{D}}\delta)^2}
+{\cal{J}}_{\rm{I}}(1-k_{\rm{I}}\delta)}{4\pi}{\bf{E}}(z),
\nonumber\\
\!\!\!\!&&\!\!\!\! z^2
=\frac{8\sqrt{{\cal{J}}^2_{\rm{H}}(1-k_{\rm{H}}\delta)^2+{\cal{D}}^2(1-k_{\rm{D}}\delta)^2}
{\cal{J}}_{\rm{I}}(1-k_{\rm{I}}\delta)}
{\left[2\sqrt{{\cal{J}}^2_{\rm{H}}(1-k_{\rm{H}}\delta)^2+{\cal{D}}^2(1-k_{\rm{D}}\delta)^2}
+{\cal{J}}_{\rm{I}}(1-k_{\rm{I}}\delta)\right]^2}. \nonumber \\
\end{eqnarray}
\end{widetext}
It is worthy to notice that the expression (\ref{4.14}) for the ground-state energy of the Heisenberg-Ising bond alternating chain 
is from the mathematical viewpoint similar to the expression (\ref{3.02}) 
determining the ground-state energy of the transverse Ising chain. 
To reduce the number of free parameters we will further assume $k_{\rm{H}}=k_{\rm{D}}\ne k_{\rm{I}}$ 
[note that the choice $k_{\rm{H}}=k_{\rm{D}}=k_{\rm{I}}$ implies that $z$ in Eq. (\ref{4.14}) becomes $\delta$-independent 
and thus, this choice leads just to a trivial case with $\varkappa=\alpha$].
Moreover, we set ${\cal{J}}_{\rm{H}}=1$, $\Delta=1$ and $k_{\rm{H}}=k_{\rm{D}}=1$,
hence leaving only three free parameters 
which control the interaction strengths and the effect of intersite distance change on the interspin interactions,
${\cal{J}}_{\rm{I}}$, ${\cal{D}}$ and $k_{\rm{I}}=k$. 
Under these assumptions, the expression (\ref{4.14}) simplifies to 
\begin{eqnarray}
\label{4.15}
e_0(\delta)&=&
-\frac{1-\delta}{8}
-\frac{2\sqrt{1+{\cal{D}}^2}(1-\delta)
+{\cal{J}}_{\rm{I}}(1-k\delta)}{4\pi}{\bf{E}}(z),
\nonumber\\
z^2
&=& \frac{8\sqrt{1+{\cal{D}}^2}{\cal{J}}_{\rm{I}}(1-\delta)(1-k\delta)}
{\left[2\sqrt{1+{\cal{D}}^2}(1-\delta)
+{\cal{J}}_{\rm{I}}(1-k\delta)\right]^2}.
\end{eqnarray}
The value of $\delta(p)$ and the inverse compressibility $\varkappa$ are given by Eqs. (\ref{3.03}) and (\ref{3.04}), respectively,
however, with $e_0(\delta)$ now given by Eq. (\ref{4.15}).

\begin{figure}[htb]
\begin{center}
\includegraphics[width=9cm]{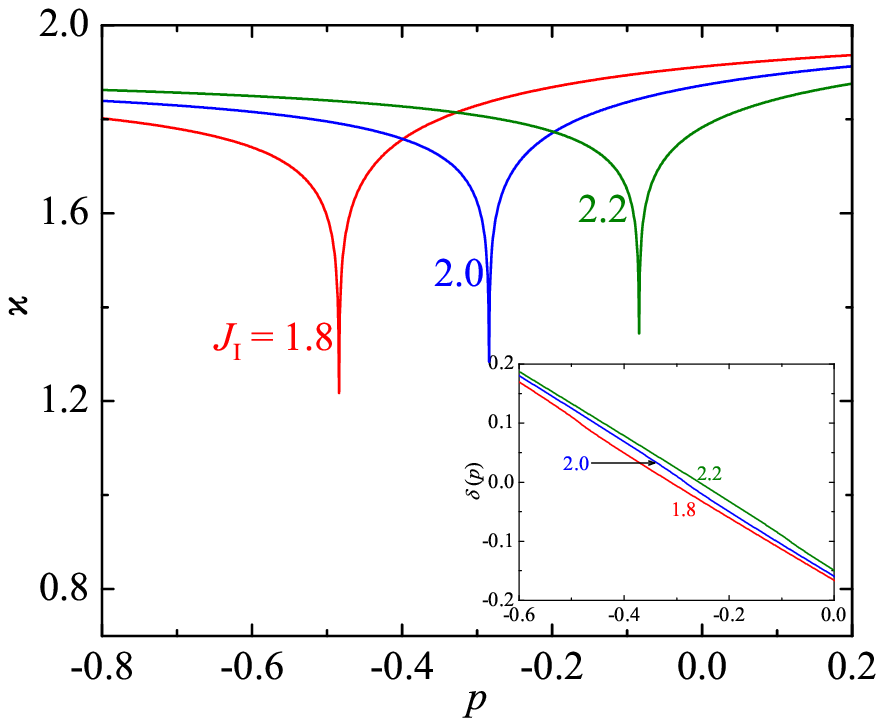}
\end{center}
\vspace{-1.2cm}
\caption{(Color online). 
Inverse compressibility $\varkappa(p)$ for the spin-1/2 Heisenberg-Ising chain (\ref{4.01}), (\ref{4.02}), (\ref{4.03})
with ${\cal{J}}_{\rm{H}}=1$, $k_{\rm{H}}=1$, $\Delta=1$, ${\cal{D}}=0$, $k_{\rm{I}}=0$, $\alpha=2$ 
and three different values of
${\cal{J}}_{\rm{I}}=1.8$ (red),
${\cal{J}}_{\rm{I}}=2$ (blue),
${\cal{J}}_{\rm{I}}=2.2$ (green).
Inset: Corresponding $\delta(p)$ versus $p$ dependences.}
\label{fig_05}
\end{figure}
\begin{figure}[htb]
\begin{center}
\includegraphics[width=9cm]{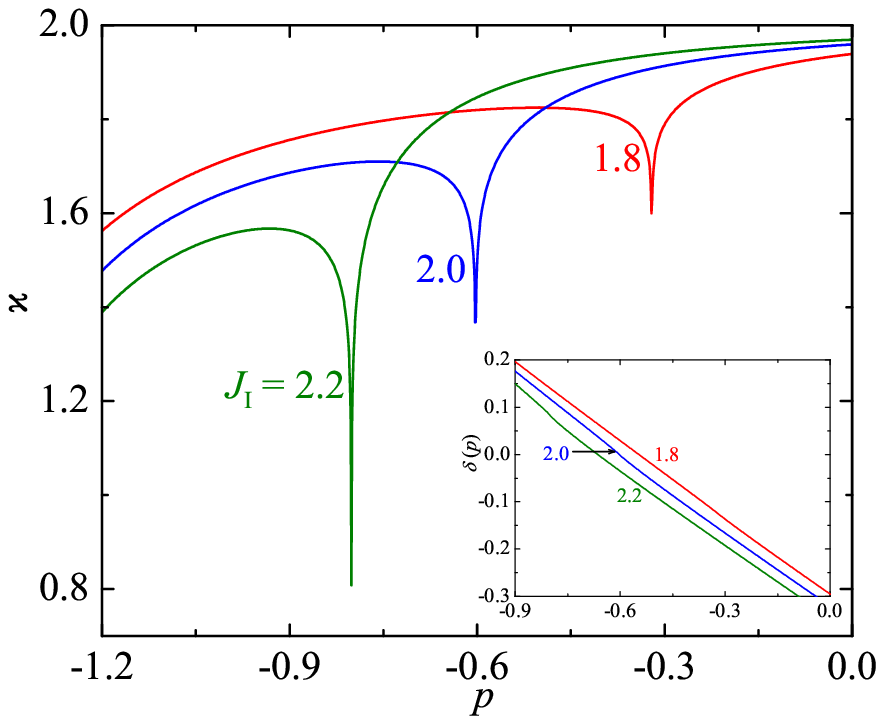}
\end{center}
\vspace{-1.2cm}
\caption{(Color online). 
Inverse compressibility $\varkappa(p)$ for the spin-1/2 Heisenberg-Ising chain (\ref{4.01}), (\ref{4.02}), (\ref{4.03})
with ${\cal{J}}_{\rm{H}}=1$, $k_{\rm{H}}=1$, $\Delta=1$, ${\cal{D}}=0$, $k_{\rm{I}}=2$, $\alpha=2$ 
and three different values of
${\cal{J}}_{\rm{I}}=1.8$ (red),
${\cal{J}}_{\rm{I}}=2$ (blue),
${\cal{J}}_{\rm{I}}=2.2$ (green).
Inset: Corresponding $\delta(p)$ versus $p$ dependences.}
\label{fig_06}
\end{figure}
\begin{figure}[htb]
\begin{center}
\includegraphics[width=9cm]{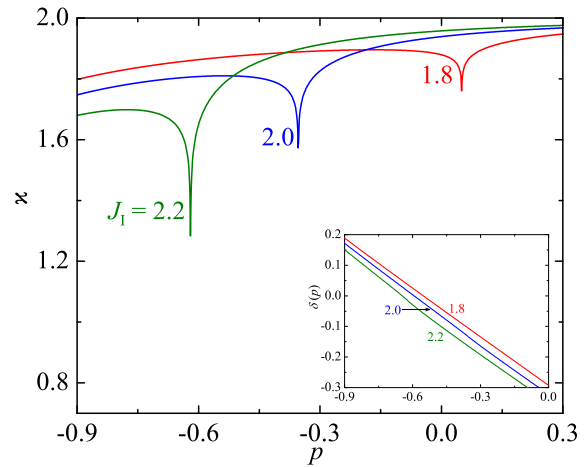}
\end{center}
\vspace{-1.2cm}
\caption{(Color online). 
Inverse compressibility $\varkappa(p)$ for the spin-1/2 Heisenberg-Ising chain (\ref{4.01}), (\ref{4.02}), (\ref{4.03})
with ${\cal{J}}_{\rm{H}}=1$, $k_{\rm{H}}=1$, $\Delta=1$, ${\cal{D}}=0.5$, $k_{\rm{D}}=1$, $k_{\rm{I}}=2$, $\alpha=2$ 
and three different values of
${\cal{J}}_{\rm{I}}=1.8$ (red),
${\cal{J}}_{\rm{I}}=2$ (blue),
${\cal{J}}_{\rm{I}}=2.2$ (green). 
Inset: Corresponding $\delta(p)$ versus $p$ dependences.}
\label{fig_07}
\end{figure}

In what follows we will discuss the effect of an applied pressure $p$ 
on elastic properties of the quantum Heisenberg-Ising bond alternating chain 
defined through the Hamiltonians (\ref{4.01}), (\ref{4.02}), (\ref{4.03}). 
It should be mentioned here that the pressure may generally induce different changes in the relevant interaction constants 
and hence, the pressure variations may be used for adjusting the interaction constants 
so as to achieve a specific condition inherent to a quantum critical point. 
In Figs.~\ref{fig_05}, \ref{fig_06}, and \ref{fig_07} 
we demonstrate the dependences of the inverse compressibility $\varkappa$ on the pressure $p$ 
along with the relevant changes of $\delta(p)$ 
at different values of ${\cal{J}}_{\rm{I}}=1.8,\;2,\;2.2$, $k=0,\;2$, and ${\cal{D}}=0,\;0.5$, 
which were obtained by solving Eqs. (\ref{3.03}), (\ref{3.04}) with $e_0(\delta)$ (\ref{4.15}) 
by subsequent numerical integration and differentiation.
It is quite clear that peculiarities in various quantities may be expected 
if the minimum of the variational enthalpy $g(T=0,p;\delta)$, 
which occurs at $\delta(p)$ given by Eqs. (\ref{3.03}) and (\ref{4.15}), 
satisfies the relation ${\cal{J}}_{\rm{I}}(1-k\delta(p))=2(1-\delta(p))\sqrt{1+{\cal{D}}^2}$,
i.e., when $\delta(p)=\delta_c$, 
\begin{eqnarray}
\label{4.16}
\delta_c = \frac{2\sqrt{1+{\cal{D}}^2}-{\cal{J}}_{\rm{I}}}{2\sqrt{1+{\cal{D}}^2}-k {\cal{J}}_{\rm{I}}}.
\end{eqnarray}
By employing Eq. (\ref{4.16}) one may subsequently calculate from Eq. (\ref{3.03}) the critical pressure  
\begin{eqnarray}
\label{4.17}
p_c= - \frac{1}{8} - \frac{2 \sqrt{1+{\cal{D}}^2}+k{\cal{J}_{\rm{I}}}}{4 \pi} 
- \alpha \frac{2\sqrt{1+{\cal{D}}^2}-{\cal{J}}_{\rm{I}}}{2\sqrt{1+{\cal{D}}^2}-k{\cal{J}}_{\rm{I}}}.
\end{eqnarray}
As one can see from the insets in Figs.~\ref{fig_05}, \ref{fig_06}, and \ref{fig_07}, 
the most robust changes in the monotonous $\delta(p)$ dependences can be observed in the neighborhood of $p=p_c$.
Note furthermore that the greater the difference between the constants $k_{\rm H}=k_{\rm D}$ and $k_{\rm I}$ is, 
the more robust are the respective deviations from a linearity in the $\delta(p)$ versus $p$ dependence. 
Because of likeness of Eq. (\ref{4.15}) and Eq. (\ref{3.02})
the behavior in the vicinity of $\delta_c$ is similar to the one treated in Sec.~\ref{sec3},
see the discussion around Eqs. (\ref{3.05}) -- (\ref{3.07}).

To gain an insight into the overall behavior, 
Fig.~\ref{fig_05} illustrates the situation 
when the pressure-induced changes of the Heisenberg interaction are greater than that of the Ising interaction 
on behalf of $k_{\rm H}>k_{\rm I}$, 
while the reverse case is displayed in Fig.~\ref{fig_06} 
when considering another possible particular case with $k_{\rm H}<k_{\rm I}$. 
Altogether, it could be concluded that the quantum spin system placed on the linear-elastic lattice 
is responsible for the deviations from Hooke's law,
i.e., the extension (or contraction) of the lattice is not linearly proportional to the applied pressure 
not because of anharmonicity of the lattice but owing to a critical behavior of interacting quantum spins placed on its sites.

Finally, let us turn our attention to the inverse compressibility $\varkappa(p)$ 
shown in the main panels in Figs.~\ref{fig_05}, \ref{fig_06}, and \ref{fig_07}. 
It can be directly observed from Figs.~\ref{fig_05} and \ref{fig_06} 
that the inverse compressibility exhibits at sufficiently low pressures a more intriguing pressure dependence 
with a substantial decline of the inverse compressibility inherent to the disordered phase 
(the regime of weak Ising interaction), 
while it always monotonically increases with rising pressure towards its maximum value $\varkappa=\alpha$ in the long-range-ordered phase 
(the regime of strong Ising interaction). 
Moreover, the quantum critical point is shifted towards higher pressures with increasing a relative strength of the Ising interaction 
in the particular case with $k_{\rm H}=k_{\rm D}>k_{\rm I}$, 
whereas the opposite trend (i.e., a shift towards lower pressures) can be observed 
in the particular case with $k_{\rm H}=k_{\rm D}<k_{\rm I}$. 
Obviously, 
the effect of the introduced Dzyaloshinskii-Moriya interaction does not fundamentally affects the physical picture we drew 
except that the relevant shift of quantum critical point due to the same change of the Ising interaction becomes much more pronounced 
as it can be seen from a comparison of Figs.~\ref{fig_06} and \ref{fig_07}. 
The most interesting finding stemming from the present study is that the inverse compressibility $\varkappa(p)$ continuously diminishes 
while approaching the critical pressure $p_c$ quite similarly as does the inverse compressibility of the transverse Ising chain, 
compare Figs.~\ref{fig_05}, \ref{fig_06}, and \ref{fig_07} with Fig.~\ref{fig_04}.
This result is taken to mean that the elastic properties of the lattice give a clear evidence 
for a quantum critical behavior of the spin-1/2 Heisenberg-Ising bond alternating chain (\ref{4.01}), (\ref{4.02}), (\ref{4.03}).

\section{Conclusions}
\label{sec5}
\setcounter{equation}{0}

Deformable spin lattices are often discussed in theoretical studies 
of, e.g., magnetic\cite{yamaguchi} or hydrogen-bonded ferro-/antiferroelectric\cite{levitskii} compounds.
However, in those studies various approximations are used which may mask detailed effect of spin-lattice coupling.
In the present work, 
we have followed rigorously how lattice elastic constant manifests peculiarities of the spin system in the ground state
driven by varying of an external field or pressure.
More precisely,
we have discussed some ground-state properties of several deformable quantum spin chains
focusing on the spin-1/2 Heisenberg-Ising bond alternating chain refined by the antisymmetric Dzyaloshinskii-Moriya interaction.
In particular,
we have followed how peculiarities in the quantum spin systems manifest themselves in the elastic properties
under simple assumption about the spin-lattice interaction.
We have found that the observed elastic constant $\varkappa$ reflects in a rather obvious manner (shows a decline)
the behavior of magnetic subsystem in the vicinity of a quantum critical point.
Although our analysis refers to the $T=0$ case, for which the studied effects are the most pronounced,
it may be generalized in a rather straightforward way for nonzero temperatures $T>0$ too.

The magnetic-field dependence of the compressibility at low enough temperatures is amenable to experimental measurements
as for instance shown in Ref.~\onlinecite{wolf1} reporting such measurements for the spin-chain material CuCCP,
or in Ref.~\onlinecite{wolf2} reporting such measurements for the two-dimensional frustrated antiferromagnet Cs$_2$CuCl$_4$.
From this perspective, it is worthwhile to remark that the elaborated analysis may be extended also for the spin-1/2 Heisenberg chain
after introducing a Hartree-Fock approximation for Jordan-Wigner fermions.

\section*{Acknowledgments}
O.~D. acknowledges financial support of the organizers of the MECO37 conference (Tatransk\'{e} Matliare, Slovakia, 18--22 March 2012)
and of the Abdus Salam International Centre for Theoretical Physics (Trieste, Italy, August 2012) where the paper was finalized.
J.~S. acknowledges financial support of Slovak Research and Development Agency provided under the grant No.~APVV-0132-11.


\begin{thebibliography}{99}
\bibitem{arbitrary}
For the Ising model in a longitudinal field 
such an analysis can be straightforwardly extended to the two- or three-dimensional case.
\end{thebibliography}
\end{document}